\documentclass[a4paper, 10pt, conference]{ieeeconf}      

\IEEEoverridecommandlockouts   

\usepackage{graphics} 
\usepackage{epsfig} 
\usepackage{mathptmx} 
\usepackage{times} 
\usepackage{amsmath} 
\usepackage{amssymb}  

\newcommand{\bdiag}{\mathop{\bf blkdiag}}
\newcommand{\diag}{\mathop{\bf diag}}
\newcommand{\ve}{\mathop{\bf vec}}
\newcommand{\reals}{{\mbox{\bf R}}}
\newcommand{\integer}{{\mbox{\bf Z}}}

\newcommand{\BEA}{\begin{eqnarray}}
\newcommand{\EEA}{\end{eqnarray}}
\newcommand{\argmin}{\mathop{\rm argmin}}

\title{\LARGE \bf
Distributed System Identification with ADMM$^*$
}

\author{Anders Hansson$^{1}$ and Michel Verhaegen$^{2}$
\thanks{*Part of the research was done while the second author was a
Visiting Professor at the Division of Automatic Control, Department of 
Electrical Engineering, Linköping University,
Sweden.  Prof. M. Verhaegen also acknowledges support from the ERC
Advanced Grant program under contract number ERC-2013-AdG339681-iCON.
Prof. A. Hansson acknowledge support from ELLIIT.}
\thanks{$^{1}$Anders Hansson is with Division of Automatic Control, 
Link\"oping University, Link\"oping, Sweden
        {\tt\small anders.g.hansson@liu.se}}%
\thanks{$^{2}$Michel Verhaegen is with Delft Center for Systems and Control,
Delft University, Delft, The Netherlands
        {\tt\small m.verhaegen@tudelft.nl}}%
}

\begin{document}
\maketitle
\thispagestyle{empty}
\pagestyle{empty}

\begin{abstract}

This paper presents identification of both network connected
systems as well as distributed systems governed by PDEs in the framework
of distributed optimization via the Alternating Direction Method of
Multipliers. This approach opens first the possibility to identify
distributed models in a global manner using all available data
sequences and second the possibility for a distributed
implementation. The latter will make the application to large scale
complex systems possible. In addition to outlining a new large
scale identification method, illustrations are shown for 
identifying both network connected systems
and discretized PDEs. 

\end{abstract}
\section{Introduction}

Control of distributed systems has recently received a renewed
interest. To just name a few examples we mention
\cite{Bamieh02,Andrea03,Massioni09,Rice11}. The interest stems from
the challenging applications that arose through the increase in
dimensionality of the systems to be controlled. 
Such increase is stimulated by various
developments, such as network communication enabling the operation of
network connected systems and/or the increase in number of
actuators and sensors for control.  An example of a network connected
systems is formation flying, \cite{Massioni11}, and an example of large scale
sensor and actuator systems is the
ongoing development of the new European Extreme Large telescope where
both the primary mirror as well as the secondary mirror are devices
with a number of sensors and actuators in the order of $10^4$ or
more, \cite{Little12}. 

A more recent development in the design of distributed controllers is
the renewed interest in distributed optimization methods from the
middle of the previous century, such as reported in
\cite{boy+par+chu+pel+eck10}. 

Despite this vast interest and despite numerous developments in the area of
distributed controller synthesis, appropriate modeling tools for deriving
the necessary models from measured data sequences are still rather
scarce. Most results are restricted to the identification of transfer
functions. In the area of identification of two dimensional (2D)
systems there is the work of \cite{Chen79} and more recently
\cite{MAli11CDC}. The last approach was developed to overcome the
difficulty in applying transfer function estimation methods that relied
on the impulse response of the system. The approach taken was to solve
the distributed identification problem as a whole using the network
topology describing the way the different systems are connected. This
approach assumes all system inputs and outputs in the network to be
available, but it avoids the problems related to the identification
of local systems in a large network topology when using only the local
input and output data. In order to derive consistent estimates with
these local identification methods, identification methods developed
for the identification under closed loop operation have to be
used, \cite{Hof13} .

In this paper we describe for the first time the identification of
distributed 2D systems and/or network connected systems in the
framework of distributed optimization methods such as the Alternating
Direction Method of Multipliers (ADMM)
\cite{boy+par+chu+pel+eck10}. We 
express distributed systems as interconnections of simple systems, and
we introduce artificial signals in order to make the resulting optimization
problem have a separable objective function. 
The use of ADMM enables us to solve the 
problem in a distributed computational manner leading to efficient
solutions for large scale problems. 

The outline of the paper as follows. In Section~II we define the
distributed identification problem. The generic framework proposed
allows us to both address problems where all input and output
measurements of systems in a given network topology are {\it known} as
well as cases with a number of the interaction variables {\it
  missing}. The latter occurs e.g in the identification of systems
governed by PDEs. In Section~III the the problem is put on a generic
form, which is suitable for making use of the ADMM algorithm in Section~IV. 
The distributed implementation is
discussed briefly in Section~V. Section~VI illustrates the methodology
for identifying ARX models connected in a feedback
topology. The application  for identifying discretized PDEs is discussed
in Section~VII.
Numerical results are summarized in
Section~VIII. Finally, in Section~IX conclusions are given together with 
directions for future research.
\section{Identification Problem}
We are interested in distributed system identification of systems that are
sparsely interconnected and where we do not measure all inputs and outputs
of the system. To fix the ideas consider systems described by 
$$\mathcal S_i(y_i,u_i,e_i,\theta_i)=0,\quad i = 1,\ldots,M,$$
where $\mathcal S_i$ is a possibly nonlinear mapping of the parameter vector
$\theta_i\in\reals^{q_i}$, the input signal vector $u_i=(u_i(1),\ldots,u_i(N))$, 
where $u_i(k)\in\reals^{m_i}$, the output signal vector 
$y_i=(y_i(1),\ldots,y_i(N))$, where $y_i(k)\in\reals^{p_i}$, and the error vector
$e_i=(e_i(1),\ldots,e_i(N))$, 
where $e_i(k)\in\reals^{p_i}$.

We assume that 
we measure the goodness of a parameter $\theta_i$ for describing 
relationship between $u_i$ and $y_i$ with a function $f_i(y_i,u_i,\theta_i)$. 
For the purpose of the remaining part of this paper we will consider 
$$f_i(y_i,u_i,\theta_i)=\|e_i\|_2^2$$
However, it should be easy to extend the result to other norms such as the 
nuclear norm. 

We will assume that the systems are interconnected according to 
\begin{eqnarray}
u(k)&=&\Gamma y(k)+Bu_0(k)\label{eqn:connection}\\
y_0(k)&=&Cy(k)\label{eqn:output}
\end{eqnarray}
where we assume that only $u_0(k)\in\reals^{m_0}$ and $y_0(k)\in\reals^{p_0}$ are 
measured. Here 
$u(k)=(u_1(k),\ldots,u_M(k))$ and $y(k)=(y_1(k),\ldots,y_M(k))$. We will also 
assume that $C$ has full row rank and that here exists a permutation matrix $P$ 
such that $CP=\begin{bmatrix}I&0\end{bmatrix}$. We also assume that 
$\begin{bmatrix}\Gamma & B\end{bmatrix}$ has only 0--1 entries and that 
it has at least one non-zero entry in each row. The remaining 
signals are just given implicitly by the above equations. Notice that we
do not assume that they are uniquely defined by these equations. However, 
we need to make the assumption that they are uniquely defined from the 
optimization problem
$$\min_{y,u,\theta}\sum_{i=1}^Mf_i(y_i,u_i,\theta_i),\quad {\rm s.\;t.}\;
(\ref{eqn:connection}-\ref{eqn:output})\quad {\rm and}\quad \theta=E\theta_0,$$
where $y=(y_1,\ldots,y_M)$, $u=(u_1,\ldots,u_M)$,
$\theta=(\theta_1,\ldots,\theta_M)$, $\theta_0\in\reals^r$ 
and $E\in\reals^{q\times r}$, with
$q=\sum_{i=1}^Mq_i$. The solution of this problem will
jointly minimize the goodness of the fit of the parameters $\theta$. We also
restrict the parameters of the different sub-models to be related to 
one another by imposing the constraint $\theta = E\theta_0$, where $E$ has
full column rank. This is typically
the case for models that come from spatial discretization of partial
differential equations. We may 
of course generalize the above problem by taking some other linear combinations
of the functions $f_i$. 
\section{Optimization Problem}
We will now cast the above problem as an optimization problem on the form
\begin{equation} \label{opt:standard}
\begin{array}{ll}
\mbox{minimize}_{(z,\theta,x,\theta_0)} & f(z,\theta) \\
\mbox{subject to} & 
z = Ax+b\\
&\theta =E\theta_0
\end{array}
\end{equation}
where $A\in\reals^{(m+p)N\times nN}$ has full column rank. 
To this end we immediately define $z_i=(y_i,u_i)$ and let 
$f(z,\theta)=\sum_{i=1}^Mf_i(y_i,u_i,\theta_i)$, where $z=(z_1,\ldots,z_M)\in
\reals^{(m+p)N}$ with $m=\sum_{i=1}^Mm_i$ and $p=\sum_{i=1}^Mp_i$.
Let $\bar y(k)$ be defined via 
$$y(k)=P\bar y(k)=\begin{bmatrix}P_1&P_2\end{bmatrix}
\begin{bmatrix}\bar y_1(k)\\\bar y_2(k)\end{bmatrix}$$
where $CP_1=I$. Because of this $\bar y_2(k)$ has dimension 
$n=\sum_{i=1}^Mp_i-p_0$. We define 
$$\bar\Gamma=\begin{bmatrix}\bar\Gamma_1&\bar\Gamma_2\end{bmatrix}=
\begin{bmatrix}\Gamma P_1&\Gamma P_2\end{bmatrix}$$
Then it holds that
$$u(k)=\bar\Gamma_1y_0(k)+\bar\Gamma_2 \bar y_2(k)+Bu_0(k)$$
We let $x(k)=\bar y_2(k)$. Then we may write
$$y(k)=P_1y_0(k)+P_2x(k)$$
We introduce
$$z(k)=\begin{bmatrix}y(k)\\u(k)\end{bmatrix}$$
We also let $z_0(k)=(y_0(k),u_0(k))$. 
From this it follows that
$$z(k)=\underbrace{\begin{bmatrix}P_2\\\bar\Gamma_2\end{bmatrix}}_{\bar A}x(k)+
\underbrace{\begin{bmatrix}P_1&0\\\bar\Gamma_1&B\end{bmatrix}}_{\bar B}
z_0(k)$$
We now introduce a permutation matrix $Q$ such that 
$$Qz(k)=\begin{bmatrix}y_1(k)\\u_1(k)\\\vdots\\y_M(k)\\u_M(k)\end{bmatrix}$$
We also let $y_0=(y_0(1),\ldots,y_0(N))$, $u_0=(u_0(1),\ldots,u_0(N))$, 
$z_0=(y_0,u_0)$, $x_i=(x_1(1),\ldots,x_i(N))$, $x=(x_1,\ldots,x_n)$,
$A = (Q\bar A)\otimes I_N$, and $B=(Q\bar B)\otimes I_N$, Then 
it holds that 
$$z=Ax+Bz_0$$
Hence $b=Bz_0$ in (\ref{opt:standard}). 
From this we realize that $A$
is a sparse matrix containing only 0--1 entries, 
and that it is a very sparse matrix if $\Gamma$ is 
sparse. Moreover, it follows that $A$ has
full column rank, since $P_2$ has full column rank. 
\section{Alternating Direction Methods of Multipliers}
We now define the augmented Lagrangian for the optimization problem in 
(\ref{opt:standard}):
\begin{eqnarray*}
L_\rho(x,\theta_0,z,\theta,\lambda,\mu)&=&f(z,\theta)+\lambda^T(z-Ax-b)\\
&+&\mu^T(\theta-E\theta_0)\\
&+&\frac{\rho}{2}\|z-Ax-b\|_2^2\\
&+&\frac{\rho}{2}\|\theta-E\theta_0\|_2^2
\end{eqnarray*}
where $\lambda\in\reals^{(m+p)N}$ and $\mu\in\reals^q$. We will from now 
on assume that $f$ is bi-convex in $z$ and $\theta$. Hence there might
be several local optima to the optimization problem. 
The Alternating Method of Multipliers (ADMM) can often successfully 
be applied to these type of problems. However, there is no guarantee for 
convergence even to local optima. The method perform alternating 
optimization steps where we need to solve $\min_{(x,\theta_0,z)}L_\rho$ for
fixed $\theta$ and $\min_{\theta}L_\rho$ for fixed $(x,\theta_0,z)$. 
Both these problems are convex, and moreover we will see that they can 
be solved by solving linear system of equations. 
There are also trivial steps in which $(\lambda,\mu)$ and possibly also 
$\rho$ are updated. 

We will now justify the bi-convexity assumption by making 
the assumption that $\mathcal S_i(y_i,u_i,e_i,\theta_i)$ 
is linear in the signals such that we may express $e_i$ as 
$$e_i=T_i(\theta_i)z_i$$
for
some matrix $T_i$ which depends linearly on $\theta_i$. 
Then
$$f_i(z_i,\theta_i)=\|T_i(\theta_i)z_i\|_2^2$$
From now on we will suppress the $\theta_i$-dependence in $T_i$.

We first consider the case of optimizing with respect to $(x,\theta_0,z)$, 
which separates into two independent optimization problems, one for
$(x,z)$ and one for $\theta_0$. 
For $\theta_0$ the augmented Lagrangian is strictly convex, and hence
the unique minimum is given by the solution of 
$$\frac{\partial L_\rho}{\partial \theta_0}=E^T\mu+\rho E^T(\theta-E\theta_0)=0$$
or equivalently of
\begin{equation}
\rho E^TE\theta_0=E^T(\mu+\rho\theta)\label{eqn:theta0update}
\end{equation}
Before we continue with the other variables we realize that if $\mu$ is 
initialized as zero, then the fact that $E^T\mu+\rho E^T(\theta-E\theta_0)=0$
together with the updated rule for $\mu$ in Table~\ref{tab:ADMM} implies that
$E^T\mu=0$, and hence (\ref{eqn:theta0update}) may be simplified to 
\begin{equation}
E^TE\theta_0=E^T\theta\label{eqn:theta0update_mod}
\end{equation}
Then for $(x,z)$ we get with similar arguments the equations:
\begin{eqnarray}
\begin{bmatrix}\frac{\partial L_\rho}{\partial z}\notag\\
\frac{\partial L_\rho}{\partial x}\end{bmatrix}&=&
\begin{bmatrix}2T^TT +\rho I&-\rho A\\-\rho A^T&\rho A^TA\end{bmatrix}
\begin{bmatrix}z\\x\end{bmatrix}\\
&+&\begin{bmatrix}\lambda -\rho b\\-A^T(\lambda -\rho b)\end{bmatrix}=0
\label{eqn:zxupdate}
\end{eqnarray}
where $T = \bdiag(T_i)$. 

We now turn our interest to solving $\min_{(\theta)}L_\rho$ for fixed 
$(x,\theta_0,z)$. We notice that the 
gradient of the Lagrangian with respect to $\theta$ is given by
\begin{eqnarray}
\frac{\partial L_\rho}{\partial \theta}&=&
\frac{\partial f}{\partial \theta}+\rho\theta+\mu-
\rho E\theta_0\notag\\
&=&2\frac{\partial e^T}{\partial\theta}T(\theta)z+\rho\theta+\mu-
\rho E\theta_0=0
\label{eqn:thetaupdate}
\end{eqnarray}
which should be zero for the optimal $\theta$. Since $T$ is linear in $\theta$
the above equation is a linear system of equations. Notice that 
$\frac{\partial e^T}{\partial\theta}$ is block diagonal, and hence the above
equations distribute nicely over $i$. We will later on for a specific model
derive more explicit equations for updating $\theta$.

We summarize the ADMM algorithm in Table~\ref{tab:ADMM}.
\begin{table}
\caption{ADMM algorithm \label{tab:ADMM}}
\begin{enumerate}
\item Set $x=0$, $\theta_0=0$, $z=b$, $\lambda=0$, $\mu=0$, $\rho=1$ and $\theta_0$ to a good guess. 
\item  Update $(x,\theta_0,z): = \argmin_{\hat x,\hat\theta_0,\hat z} L_\rho(\hat x,\hat\theta_0,\hat z,\theta,\lambda)$. 
\item Update $\theta:= \argmin_{,\hat\theta} L_\rho(x,\theta_0,z,\hat\theta,\lambda)$.
\item Update $(\lambda,\mu) := (\lambda + \rho (z-Ax-b),\mu+\rho(\theta-E\theta_0)$.
\item Terminate if
  $\|r_\mathrm{p}\|_2 \leq \epsilon_\mathrm{p}$
and $\|r_\mathrm{d}\|_2 \leq \epsilon_\mathrm{d}$  
 (see~(\ref{e-rp})--(\ref{e-ed})).  
 Otherwise, go to step 2. 
\end{enumerate}
\end{table}
The residuals and tolerances in the stopping criterion 
in step~5 are defined as follows \cite{boy+par+chu+pel+eck10}:
\BEA
 r_\mathrm{p} & = & (z-Ax-b,\theta-E\theta_0)\label{e-rp}\\
 r_\mathrm{d} & = & \rho (A^T (z_\mathrm{prev} - z),E^T(\theta_\mathrm{prev}-\theta)) \label{e-rd} \\
 \epsilon_\mathrm{p} & = & \sqrt{(m+p)N+q} \, \epsilon_\mathrm{abs} \\
 & + &
\epsilon_\mathrm{rel} \max \{\|(Ax,E\theta_0)\|_2, \|(z,\theta)\|_2, 
\| b\|_2 \} \label{e-er} \\
 \epsilon_\mathrm{d} & = & \sqrt{nN+r} \epsilon_\mathrm{abs} +
\epsilon_\mathrm{rel} \, \| (A^T\lambda,E^T\mu) \|_2, \label{e-ed}
\EEA
Typical values for the relative and absolute tolerances are 
$\epsilon_\mathrm{rel} = 10^{-3}$
and $\epsilon_\mathrm{abs} = 10^{-6}$.  
The vectors $z_\mathrm{prev}$ and $\theta_\mathrm{prev}$ 
in~(\ref{e-rd}) are the values of $z$ and $\theta$ in 
the previous iteration.

Instead of a using a fixed penalty parameter $\rho$, 
one can vary $\rho$ to improve the speed of convergence.
An example of such a scheme is to adapt $\rho$ at the
end of each ADMM iteration as follows~\cite{boy+par+chu+pel+eck10}
\[
 \rho := \left\{ 
  \begin{array}{ll}
   \tau \rho \quad & \|r_\mathrm{p}\|_2 > \mu \|r_\mathrm{d}\|_2 \\
   \rho/\tau \quad & \|r_\mathrm{d}\|_2 > \mu \|r_\mathrm{p}\|_2 \\
   \rho \quad & \mbox{otherwise.}
  \end{array} \right.
\]
This scheme depends on parameters $\mu >1$, $\tau>1$ 
(for example, $\mu = 10$ and $\tau=2$).
\section{Distributed Implementation}
We have so far seen that the equations for updating $\theta$ in 
(\ref{eqn:thetaupdate}) can be carried out distributively over $i=1,\ldots,M$
by solving
\begin{eqnarray*}
\frac{\partial L_\rho}{\partial \theta_i}&=&
\frac{\partial f_i}{\partial \theta_i}+\rho\theta_i+\mu_i-
\rho (E\theta_0)_i\\
&=& 2\frac{\partial e_i^T}{\partial\theta_i}T_i(\theta_i)z_i+\rho\theta_i+\mu_i-
\rho (E\theta_0)_i=0
\end{eqnarray*}
because $\frac{\partial e^T}{\partial\theta}$ and $T(\theta)$ are block
diagonal. In the right hand side we are however interested in explaining
the term $(E\theta_0)_i$ further. 
It is not uncommon that $E$ is an incidence matrix
of zeros and ones describing what component of $\theta_0$ is related to each
component in $\theta$. We write
$$E = \begin{bmatrix}E_1\\\vdots\\E_M\end{bmatrix}$$
where the partitioning is done conformable with the partitioning of $\theta$. 
In a graph setting we consider each component of
$\theta_0$ to be represented by its index in the vertex set 
$\mathcal V_0=\{1,\ldots,q_0\}\subset\integer$ and each component of
$\theta_i$ to be represented by its index in the vertex set 
$\mathcal V_i=\{1,\ldots,q_i\}\subset\integer$. The $i$th 
graph has a directed edge 
$e\in \mathcal V_0\times \mathcal V_i$
if and only $(E_i)_e=1$. We denote the set of all edges of the graph by
$\mathcal E_{\theta_i}$. 
It then follows that we may write
$$2\frac{\partial e_i^T}{\partial\theta_i}T_i(\theta_i)z_i+\rho\theta_i+\mu_i-
\rho \bar\theta_i=0$$
where $\bar\theta_{i,k}=\theta_{0,j}$ if $(j,k)\in\mathcal E_{\theta_i}$ and
zero otherwise. 
Hence for each $i$ information is needed only from the components of $\theta_0$
that are defining $\theta_i$. 

We will now discuss how also (\ref{eqn:theta0update_mod}) and
(\ref{eqn:zxupdate}) distribute over $i$. 
First we consider 
(\ref{eqn:theta0update_mod}).
The out degree $d_{0,i}(j)$ of a vertex $j\in \mathcal V_0$ is 
the number of edges that emerges from it in graph $\mathcal E_{\theta_i}$. 
It follows that 
$$E^TE=\diag_j(d_0(j))$$
where $d_0(j)=\sum_{i=1}^Md_{0,i}(j)$. 
We now realize that we can updated each component in $\theta_0$ using the
formula
$$\theta_{0,j}=\frac{1}{d_0(j)}\sum_{(j,k)\in\mathcal E_{\theta_i}}\theta_{i,k},
\quad j \in\mathcal V_0$$
We see that we only sum over those components of $\theta$ which are 
defined by $\theta_{0,j}$, and that the computations can be performed 
locally for each component of $\theta_0$.

We now consider (\ref{eqn:zxupdate}). We notice 
that we can first solve 
\begin{equation}
\rho A^T\left(I-\rho(2T^TT+\rho I)^{-1}\right)Ax=
-A^T\left(I-\rho(2T^TT+\rho I)^{-1}\right)r
\label{eqn:xupdate}
\end{equation}
with respect to $x$, where $r=\lambda -\rho b$. Then we can solve
\begin{equation}
(2T_i^TT_i+\rho I)z_i=\rho (Ax)_i-r_i
\label{eqn:zupdate}
\end{equation}
with respect to $z_i$ for $i=1,\ldots,M$. The latter equation clearly 
distributes over $i$ for the left hand side, and for the right hand side
we are interested in what information about $x$
that is needed for each block $i$, i.e. what $(Ax)_i$ is. We remember that
$A=(Q\bar A)\otimes I_N$, that $\bar A$ is a zero one matrix, and that $Q$ is a 
permutation matrix. Hence $A$ is also a zero one matrix. We let 
$\tilde A=Q\bar A$, and we partition it as
$$\tilde A=\begin{bmatrix}\tilde A_1\\\vdots\\\tilde A_M\end{bmatrix}$$
where the partitioning is done conformable with $z$. Then $(Ax)_i=(\tilde A_i
\otimes I_N)x$, and hence we may rewrite (\ref{eqn:zupdate}) as
$$(2T_i^TT_i+\rho I)z_i=\rho(\tilde A_i\otimes I_N)x -r_i,\quad i=1,\ldots,M$$
Hence we are able to update each $z_i$ locally with information only from 
those components of $x$ which are used to explain $z_i$. 

We now turn our interest to-wards (\ref{eqn:xupdate}) and define
$X_i=I-\rho(2T_i^TT_i+\rho I)^{-1}$ and $X=\bdiag X_i$. We then realize that
$A^TXA=\sum_{i=1}^M(\tilde A_i^TX_i\tilde A_i)\otimes I_N$, and hence
$$x=-\frac{1}{\rho}\left\lbrace\left[
\left(\sum_{i=1}^M\tilde A_i^TX_i\tilde A_i\right)^{-1}\sum_{i=1}^M\tilde A_i^TX_i
\right]\otimes I_N\right\rbrace r$$ 
We see that we need global information in order to carry out the update of
$x$. However, we also realize that the matrix that needs to be inverted only
has dimension $n$, which is typically low.  
\section{Feedback Connection of ARX-Models}
In this section we will give a description of a simple feedback connection of
three ARX models:
\begin{eqnarray}
y_i(k)&+&a_{i,1}y_i(k-1)+a_{i,2}y(k-2)\\
&+&b_{i,1}u(k-1)+b_{i,2}u(k-2)=e_i(k)
\end{eqnarray}
where $k=1,\ldots,N$ and $i=1,2,3$. We let
$\theta_i=(a_i,b_i)\in\reals^{4}$, and we define $\theta_0$ such that we may
take $E=I$, i.e. the parameters of the models are not constrained in any
way. 
The interconnection matrices are given by 
$$\Gamma = \begin{bmatrix}0 & 0 & -1\\1 & 0 & 0 \\ 0 & 1 & 0\end{bmatrix};
\quad B=\begin{bmatrix}1\\0\\0\end{bmatrix}$$
Moreover we measure all outputs, i.e. $C=I$. We may write 
$$e_i=\Phi_i\theta_i+y_i$$
where $\Phi_i=\begin{bmatrix}Sy_i&S^2y_i&Su_i&S^2u_i\end{bmatrix}$, where $S$
is a shift matrix. Hence (\ref{eqn:thetaupdate}) may be equivalently written as
$$\left(\Phi^T\Phi+\rho I\right)\theta=
\rho E\theta_0 -\mu-2\Phi y$$
where $\Phi=\bdiag \Phi_i$. The distributed version is
$$\left(\Phi_i^T\Phi_i+\rho I\right)\theta_i=
\rho \bar\theta_i -\mu_i-2\Phi_i y_i,\quad i=1,\ldots,M$$
We remark that for this example the dimension $n$ of the $x$-variable is zero. 
\section{Discretized Partial Differential Equation}
We will also consider a model that comes from a spatial discretization of 
a partial differential equation, which is defined as
$$y_i(k)+(a_i)^T\begin{bmatrix}y_i(k-1)\\y_i(k-2)\end{bmatrix}=
(b_i)^Tu_i(k)+e_i(k),\quad i = 1,\ldots,M$$
where $u_1(k),\;u_M(k)\in\reals^3$, $u_2(k),\;u_{M-1}(k)\in\reals^4$, and
$u_i(k)\in\reals^5$ for $i=3,\ldots,M-2$, and where $y_i(k),\;e_i(k)\in\reals$. 
The dimensions of $a_i$ and
$b_i$ are compatible with the signal dimensions. 
The inputs are partially feedbacks from the neighboring 
systems according to 
\begin{eqnarray}
u_1(k)&=&\begin{bmatrix}u_{0,1}(k)\\y_2(k)\\y_3(k)\end{bmatrix}\\
u_2(k)&=&\begin{bmatrix}y_1(k)\\u_{0,2}(k)\\y_3(k)\\y_4(k)\end{bmatrix}\\
u_i(k)&=&\begin{bmatrix}y_{i-2}(k)\\y_{i-1}(k)\\u_{0,i}(k)\\y_{i+1}(k)\\y_{i+2}(k)\end{bmatrix},\quad i=3,\ldots,M-2\\
u_{M-1}(k)&=&\begin{bmatrix}y_{M-3}(k)\\y_{M-2}(k)\\u_{0,M-1}(k)\\y_{M}(k)\end{bmatrix}\\
u_{M}(k)&=&\begin{bmatrix}y_{M-2}(k)\\y_{M-1}(k)\\u_{0,M}(k)\end{bmatrix}
\end{eqnarray}
where $u_{0,i}(k)$ are measured inputs. This
defines the matrices $\Gamma$ and $B$. Moreover we measure every second 
output $y_i(k)$, i.e. 
$$C=\begin{bmatrix}e_1^T\\e_3^T\\\vdots\\e_{M-2}^T\\e_M^T\end{bmatrix}$$
where $e_i$ is the $i$th unit vector with abuse of notation. We will also
assume that $M\geq 5$ and that $M$ is an odd integer. We let
$\theta_0=(a_0,b_0)\in\reals^5$, $\theta_i=(a_i,b_i)\in\reals^{2+m_i}$. 
We then define the constraints $a_i=a_0$ and 
\begin{eqnarray}
b_1&=&b_0\\
b_2(k)&=&\begin{bmatrix}e_2^T\\I_3\end{bmatrix}b_0\\
b_i(k)&=&\begin{bmatrix}e_3^T\\e_2^T\\I_3\end{bmatrix}b_0,\quad i=3,\ldots,M-2\\
b_{M-1}(k)&=&\begin{bmatrix}e_3^T\\e_2^T\\e_1^T\\e_2^T\end{bmatrix}b_0\\
b_{M}(k)&=&\begin{bmatrix}e_2^T\\e_1^T\\e_2^T\end{bmatrix}b_0
\end{eqnarray}
where $e_i$ is the $i$th unit vector in $\reals^3$. This defines $E$, and
the overall model. We now define
$$\Phi_i=\begin{bmatrix}Sy_i&S^2y_i&-U^T\end{bmatrix}$$
where $S$ is a shift matrix of compatible dimension and where $U$ is such that
$S^{m_i}u_i = \ve (U)$ with $\ve$ being the vectorization operator. Here 
$S$ has different dimension depending on where it appears. Then
$$e_i=\Phi_i\theta_i+y_i$$
and hence (\ref{eqn:thetaupdate}) may be equivalently written as
$$\left(\Phi^T\Phi+\rho I\right)\theta=
\rho E\theta_0 -\mu-2\Phi y$$
where $\Phi=\bdiag \Phi_i$. The distributed version is
$$\left(\Phi_i^T\Phi_i+\rho I\right)\theta_i=
\rho \bar\theta_i -\mu_i-2\Phi_i y_i,\quad i=1,\ldots,M$$
\section{Numerical Experiments}
All implementations have been carried out in
MATLAB R2013b. The computations have been run on an Intel Core i5 CPU M 250 
4 GHz with 4 GB of RAM.
\subsection{Feedback Connection of ARX-Models}
All ARX models have been defined as $a_i=(-1.5,0.7)$ and $b_i=(-0.1,0.1)$ for
$i=1,2,3$. The input $u_0$
has been taken as a sequence of independent $\pm 1$-variables. The
error vector $e$ has been generated from a zero mean normal density function
with standard deviation $\sigma=1$. 
Then the closed loop signals have been computed
from the equations
\begin{eqnarray}
&&\left(\bdiag(T_{y,i})+\bdiag(T_{u,i})(\Gamma\otimes I_N)\right)y=\\
&&\left(-\bdiag(T_{u,i})(B\otimes I_N)\right)u_0+e\\
&&u = (\Gamma\otimes I_N)y + (B\otimes I_N)u_0\\
&&y_0 = (C\otimes I_N)y
\end{eqnarray}
The value of $N$
has been 300. We have used the default settings
for the ADMM algorithm as detailed above.
The initial guess for $\theta_0$ was the zero vector.

We repeated the optimization 100 times.
The mean value of the estimated parameters 
were
\begin{eqnarray*}
m_{\theta_1}&=&\begin{bmatrix}-1.4988    &0.7013   &-0.0964    &0.0965 \end{bmatrix}^T\\
m_{\theta_2}&=&\begin{bmatrix}-1.4934    &0.6923   &-0.1068   & 0.1071\end{bmatrix}^T\\
m_{\theta_3}&=&\begin{bmatrix}-1.4897    &0.6896   &-0.1105    &0.1084\end{bmatrix}^T
\end{eqnarray*}
with standard deviations
\begin{eqnarray*}
\sigma_{\theta_1}&=&\begin{bmatrix}0.0371    &0.0385    &0.0321    &0.0315\end{bmatrix}^T\\
\sigma_{\theta_2}&=&\begin{bmatrix}0.0457    &0.0473    &0.0342    &0.0349\end{bmatrix}^T\\
\sigma_{\theta_3}&=&\begin{bmatrix}0.0435    &0.0408    &0.0476    &0.0473\end{bmatrix}^T
\end{eqnarray*} 
We see that the model parameters are estimated accurately.
\subsection{Discretized Partial Differential Equation}
The dynamical system considered has been $a_0=(0.7,0.9)$ and 
$b_0=(0.5,-0.5,0.5)$.
The input $u_0$
has been taken as a sequence of independent $\pm 1$-variables. The
error vector $e$ has been generated from a zero mean normal density function
with standard deviation $\sigma=1$. 
Then the closed loop signals have been computed in the same way as for the 
previous example.
The value of $N$
has been 100 and the value of $M$ has been 15. We have used the default settings
for the ADMM algorithm as detailed above except for 
$\epsilon_\mathrm{rel} = 10^{-1}$
and $\epsilon_\mathrm{abs} = 10^{-4}$, which provided good enough solutions.
The initial guess for $\theta_0$ was the true value of its components
perturbed with 
a value drawn from a zero mean normal density with standard deviation 0.1. 

We repeated the optimization 10 times and we report in Table~\ref{tab:results}
computational time, and the number of iterations in the ADMM algorithm
for the different runs. 
\begin{table*}[htbp]
\centering
\caption{Iterations and Time \label{tab:results}}
\begin{tabular}{c|r r r r r r r r r r}
Run nr & 1 & 2 & 3 & 4 & 5 & 6 & 7 & 8 & 9 & 10\\
\hline
Iterations &177 &107 &77 &135 &34 &84 &306 &95 &177 &105\\
Time (s) & 515.9 &306.2 &219.4 &406.6 &106.8 &246.3 &3164.8 &262.5 &496.9 &305.9
\end{tabular}
\end{table*}
The mean value of the estimated parameters 
were
$$m_{\theta_0}=\begin{bmatrix}0.7017 &0.8950 &0.4958 &-0.4966 &0.4957\end{bmatrix}^T$$ 
with standard deviation
$$\sigma_{\theta_0}=\begin{bmatrix}0.0075 &0.0110 &0.0212 &0.0089 &0.0086\end{bmatrix}^T$$ 
It is seen that the proposed algorithm computes good estimates of the 
true parameters in reasonable time. It should be stressed that we have not
made use of parallel or distributed implementations. Hence the computational
times should be possible to decrease significantly. It should also
be noted that 
the our results relay on a good initial guess of $\theta_0$. 
\section{Summary}
To summarize it looks like it should be possible to solve identification
problems of interconnected systems where we do not measure all 
input or output signals in a distributed way. 
An open question is how much need to 
be measured to have a unique solution. Also can this framework be used to 
solve identification problems for state space descriptions when one
impose structure on the system matrices? Our framework addresses as a special
case distributed estimation of signals by assuming that $\theta$ is known. 
We admit that in case no good
guess of the true parameters are available to initialize the ADMM algorithm,
it may fail to find the global optimal solution. It may instead be trapped
in a local minimum. Future research will investigate the possibility to use
continuation methods to remedy this flaw. 
\bibliography{myref}
\bibliographystyle{plain}
\end{document}